%% file: dsn-rr.tex
\documentclass[a4paper,  10pt]{article}
\usepackage{RRA4}
\usepackage{hyperref}

\usepackage[latin1]{inputenc}
\usepackage{alltt}
\usepackage{epsfig}
\usepackage{./frame}
\usepackage{url}
\DeclareUrlCommand\fname{\urlstyle{tt}}

\input{macros.tex}

\RRdate{F\'evrier 2007}
\RRauthor{% les auteurs
Laurent Burgy 
\and
Laurent R\'eveill\`ere
\and
Julia Lawall
\and 
Gilles Muller
}
\authorhead{Burgy, R\'eveill\`ere, Lawall \& Muller}
\RRtitle{Une approche langage pour am\'eliorer la robustesse de l'impl\'ementation de protocoles r\'eseaux applicatifs}
\RRetitle{A Language-Based Approach for Improving the Robustness of Network Application Protocol Implementations}
\titlehead{A Language-Based Approach for Improving the Robustness of Network Application Protocol Implementations}
\RRresume{
Pour \^etre s\^ur et robuste, le fonctionnement d'un r\'eseau doit reposer sur des impl\'ementations d'applications sans faille. Les protocoles r\'eseau \'etant de plus en plus complexes, \'ecrire manuellement le code qui prend en charge leurs messages devient de plus en plus difficile et sujet \`a erreurs.

Dans ce papier, nous pr\'esent un langage m\'etier, Zebu, pour d\'ecrire le format des messages d'un protocole r\'eseau et les contraintes de traitement associ\'ees. D'une sp\'ecification Zebu, un compilateur g\'en\`ere automatiquement des talons \`a utiliser par une application pour l'analyze grammaticale de messages r\'eseau. Zebu is simple d'usage, utilisant les m\^emes notations que celles utilis\'ees dans les RFCs pour d\'ecrire les grammaires de protocoles. Zebu est efficace, l'implantation m\'emoire \'etant calqu\'ee sur les besoins de l'application et les fragments du message pouvant \^etre trait\'es \`a la demande. Enfin, les applications bas\'ees sur Zebu sont robustes, le compilateur v\'erifiant la consistence de la sp\'ecification et les talons g\'en\'er\'es \'etant incluant la validation de la structure du message. En utilisant une analyze de mutation dans le contexte de SIP et RTSP, nous montrons que Zebu am\'eliore de mani\`ere significative la robustesse des applications.
}
\RRabstract{ The secure and robust functioning of a network relies on the
  defect-free implementation of network applications. As network
  protocols have become increasingly complex, however, hand-writing
  network message processing code has become increasingly error-prone.

  In this paper, we present a domain-specific language, Zebu, for
  describing protocol message formats and related processing
  constraints. From a Zebu specification, a compiler automatically
  generates stubs to be used by an application to parse network
  messages. Zebu is easy to use, as it builds on notations used in RFCs to
  describe protocol grammars. Zebu is also efficient, as the memory usage
  is tailored to application needs and message fragments can be specified
  to be processed on demand. Finally, Zebu-based applications are robust,
  as the Zebu compiler automatically checks specification consistency and
  generates parsing stubs that include validation of the message structure.
  Using a mutation analysis in the context of SIP and RTSP, we show that
  Zebu significantly improves application robustness.
}
\RRmotcle{DSL, network protocols, mutation analysis}
\RRkeyword{Langage m\'etier, protocoles r\'eseau, analyze de mutation}
\RRprojets{Phoenix et Obasco}
\RRtheme{\THCom} % cas de 5 themes
\URFuturs % pour ceux qui sont dans le virtuel
\begin{document}
\makeRR   % cas d'un rapport de recherche
%% \makeRT % cas d'un rapport technique.
%% a partir d'ici, chacun fait comme il le souhaite
%\tableofcontents
%\newpage
%\onecolumn
\tableofcontents
%\twocolumn

%------------------------------------------------------------------------- 
% Section: Introcuction
\input{introduction.tex}

%------------------------------------------------------------------------- 
% Section: Context
\input{issues.tex}

%------------------------------------------------------------------------- 
% Section: Robust Message Handling Layer Development with Zebu
\input{robustdev.tex}

%------------------------------------------------------------------------- 
% Section: Experiments
\input{experiments.tex}

%------------------------------------------------------------------------- 
% Section: Related works
\input{related.tex}

%------------------------------------------------------------------------- 
% Section: Conclusion and future directions
\input{conclusion.tex}

\bibliographystyle{abbrv}
\bibliography{related,paper,phoenix}

\end{document}

%% file: macros.tex
\let\codesize\scriptsize

\def\nlineskip#1{\vskip#1\baselineskip}

\def\{{{\char123}}
\def\}{{\char125}}
\def\_{{\char95}}

\newcommand{\mtt}[1]{\mbox{\texttt{{#1}}}}

\def\doframeit#1{\vbox{%
  \hrule height\fboxrule
    \hbox{%
      \vrule width\fboxrule \kern\fboxsep
      \vbox{\kern\fboxsep #1\kern\fboxsep }%
      \kern\fboxsep \vrule width\fboxrule }%
    \hrule height\fboxrule }}

\def\frameit{\smallskip \advance \linewidth by -7.5pt \setbox0=\vbox \bgroup
\strut \ignorespaces }

\def\endframeit{\ifhmode \par \nointerlineskip \fi \egroup
\doframeit{\box0}}

\def\@addmarginpar{\@next\@marbox\@currlist{\@cons\@freelist\@marbox
    \@cons\@freelist\@currbox}\@latexbug\@tempcnta\@ne
    \@tempcnta\m@ne % always to left margin <---
    \ifnum\@tempcnta <\z@ \global\setbox\@marbox\box\@currbox \fi
    \@tempdima\@mparbottom \advance\@tempdima -\@pageht
       \advance\@tempdima\ht\@marbox \ifdim\@tempdima >\z@
       \@warning{Marginpar on page \thepage\space moved}\else\@tempdima\z@ \fi
    \global\@mparbottom\@pageht \global\advance\@mparbottom\@tempdima
       \global\advance\@mparbottom\dp\@marbox
       \global\advance\@mparbottom\marginparpush
    \advance\@tempdima -\ht\@marbox
    \global\ht\@marbox\z@ \global\dp\@marbox\z@
    \vskip -\@pagedp \vskip\@tempdima\nointerlineskip
    \hbox to\columnwidth
      {\ifnum \@tempcnta >\z@
          \hskip\columnwidth \hskip\marginparsep
        \else \hskip -\marginparsep \hskip -\marginparwidth \fi
       \box\@marbox \hss}
    \vskip -\@tempdima
    \nointerlineskip
    \hbox{\vrule \@height\z@ \@width\z@ \@depth\@pagedp}}

%% file: introduction.tex
\section{Introduction}\label{sec:intro}

In the Internet era, many applications, ranging from instant messaging
clients and multimedia players to HTTP servers and proxies, involve
processing network protocol messages.  A key part of this processing is to
parse messages as they are received from the network.  As message parsing
represents the front line of interaction between the application and the
outside world, the correctness of the parser is critical; any bugs can
leave the application open to attack \cite{Qie:osdi02}.  In the context of
in-network application such as proxies, where achieving high throughput is
essential, parsing must also be efficient.

Implementing a correct and efficient network protocol message parser,
however, is a difficult task.  The syntax of network protocol messages is
typically specified in a RFC (\textit{Request for Comments}) using a
variant of BNF known as ABNF (\textit{Augmented BNF}) \cite{rfc2234}.  Such
a specification amounts to a state machine, which for efficiency is often
implemented in an unstructured way using gotos.  The resulting code is thus
error-prone and difficult to maintain.  Furthermore, some kinds of message
processing may not use all fragments of the message.  For example, a router
normally only uses the header fields that describe the message destination,
and ignores the header fields that describe properties of the message
body~\cite{kuthan:ser-03}.  It is thus desirable, for efficiency, to defer
the parsing of certain message fragments to when their values are actually
used.  In this case, complex parsing code may end up scattered throughout
the application.

In the programming languages community, parsers have long been constructed
using automated parser generators such as \texttt{yacc} \cite{yacc}.
Nevertheless, such tools are not suitable for generating parsers for
network protocol messages, as the grammars provided in RFCs are often not
context free, and such tools provide no support for deferring the parsing
of some message fragments.  Thus, parsers for network protocol messages
have traditionally been implemented by hand.  This situation, however, is
becoming increasingly impractical, given the variety and complexity of
protocols that are continually being developed.  For example, the Gaim
instant messaging client parses more than 10 different instant messaging
protocols \cite{gaim}.  The message grammar in the IMAP RFC is about 500
lines of ABNF, and includes external references to others RFCs.  SIP ({\em
Session Initiation Protocol})~\cite{sip_rfc3261}, which is mainly used in
telephony over IP, has a multitude of variants and extensions, implying
that SIP parsers must be tolerant of minor variations in the message
structure and be extensible.  Incorrect or inefficient parsing makes the
application vulnerable to denial of service attacks, as illustrated by the
``leading slash'' vulnerability found in the Flash HTTP Web server
\cite{Qie:osdi02}.  In our experiments (Section
\ref{experiments}), we have crashed the widely used SER
parser~\cite{kuthan:ser-03} for SIP via a stream of 2416 incorrect
messages, sent within 17 seconds.

To address the growing complexity of network protocol messages and the
inadequacy of standard tools, some parser generators have recently been
developed that specifically target the kinds of complex data layouts found
in network protocol messages.  These tools include
DATASCRIPT~\cite{Back:gpce02} and
PacketTypes~\cite{DBLP:conf/sigcomm/McCannC00} for binary protocols, and
PADS~\cite{DBLP:conf/pldi/FisherG05}, GAPA~\cite{gapa} and
binpac~\cite{Pang:im06} for both binary and text-based protocols.  However,
none of these approaches accepts ABNF as the input language, and thus, the
RFC specification must be translated to another formalism, which is tedious
and error prone.  Furthermore, such approaches have mainly targeted
application protocol analyzers, which parse a fixed portion of the message
and then proceed to some analysis phase. Thus, they do not provide
fine-grained control over the time when parsing occurs.  While these
approaches relieve some of the burden of implementing a network protocol
message parser, there still remains a gap between these tools and the needs of
applications.

We propose to directly address the issues of correctness and efficiency at
the parser generator level.  To this end, we present a domain-specific
language, Zebu, for describing HTTP-like text-based protocol message
formats and related processing constraints.  Zebu is an extension of ABNF,
implying that the programmer can simply copy a network protocol message
grammar from an RFC to begin developing a parser.  Zebu extends ABNF with
annotations indicating which message fields should be stored in data
structures, and other semantic information, such as the type of the value
expressed by a field, constraints on the range of its value, and whether
certain fields are mandatory or optional.  Fields can additionally be
declared as {\tt lazy}, which gives control over the time when the parsing
of a field occurs.  A Zebu specification is then processed by a compiler
that generates stubs to be used by an application to process network
messages.  Based on the annotations, the Zebu compiler implements
domain-specific optimizations to reduce the memory usage of a Zebu based
application.  Besides efficiency, Zebu also addresses robustness, as the
compiler performs many consistency checks, and generates parsing stubs that
validate the message structure.

\paragraph{This paper}

In this paper, we present the Zebu language and an assessment of its
performance and robustness in the context of the SIP and RTSP (Real Time
Streaming Protocol \cite{rtsp}) network protocols.  Our contributions are
as follows:
\begin{itemize}

\item We introduce a declarative language, named Zebu, for describing
  protocol message formats and related processing constraints. Zebu builds
  on the ABNF notation typically used in RFCs to describe protocol grammars.

\item We have defined a test methodology based on a mutation analysis for
  evaluating the robustness improvement induced by Zebu.

\item We have applied our test methodology existing and Zebu-based SIP and
  RTSP parsers. While the Zebu-based parsers reject 100\% of the invalid
  mutated messages, none of the existing parsers that we have tested
  detects more than about 25\% of the injected mutants.

\item Finally, we show that the added safety and robustness provided by
  Zebu does not significantly impact performance.  Indeed, our performance
  evaluation shows that a Zebu-based parser can be as efficient on average as a
  hand-crafted one.

\end{itemize}

The rest of this paper is organized as follows. Section~\ref{issues}
discusses specific characteristics of network protocol message parsing
code, illustrating its inherent complexity.  Section~\ref{language}
introduces the Zebu language, and describes the verification of
specifications and the generation of parsing stub functions.
Section~\ref{experiments} assesses the robustness and performance of
Zebu-based parsers.  Section~\ref{rel} described related work and
Section~\ref{conclusion} concludes.

%% file: issues.tex
\section{Issues in developing network protocol \newline parsers} 
\label{issues}

To illustrate the growing complexity of network protocol messages and the
inadequacy of existing approaches to creating the associated parsers, we
consider the SIP protocol \cite{sip_rfc3261}.  The SIP message syntax is
similar to that of other recent text-based protocols such as HTTP and RTSP.
A SIP message begins with a line indicating whether the message is a
request (including a protocol method name) or a response (including a
return code). A sequence of required and optional headers then follows.
Finally, a SIP message can include a body containing the payload.  Widely
used SIP parsers include that of the SIP Express Router
(SER)~\cite{kuthan:ser-03} and the oSIP library~\cite{osip} used {\em
e.g.}, in the open PBX Asterisk~\cite{asterisk}.  Both parsers are
hand-written.

%% SIP is a text-based protocol
%% that is mainly used in telephony over IP to initiate and terminate voice
%% sessions.

We first present an extract of the ABNF specification of the SIP message
grammar, and then describe the difficulty of hand-writing the corresponding
parser.  We next consider to what extent these difficulties are
addressed by existing parser generation tools, and describe the issues
involved in integrating a parser with a network application.

\subsection{ABNF formalism}
\label{issues_abnf}

An extract of the ABNF specification of the SIP message grammar is shown in
Figure~\ref{fig:rfc3261}.  Lines 1 to 8 define the structure of a request
line, which appears at the beginning of a message, and lines 10 to 13
define the structure of the CSeq header field, which is used to identify
the collection of messages making up a single transaction.

\begin{figure}[t]
  \centering\scriptsize
  \begin{alltt}
 Request-Line = Method SP Request-URI SP SIP-Version CRLF \n{1}
 Method       = INVITEm / ACKm / OPTIONSm / BYEm          \n{2}
                / CANCELm / REGISTERm                     \n{3}
                / extension-method                        \n{4}
 INVITEm      = %x49.4E.56.49.54.45 ; INVITE in caps      \n{5}
 Request-URI  = SIP-URI / SIPS-URI / absoluteURI          \n{6}
 SIP-Version    =  "SIP" "/" 1*DIGIT "." 1*DIGIT          \n{7}
 extension-method = token                                 \n{8}
 \textit{[...]}
 CSeq = "CSeq" HCOLON 1*DIGIT LWS Method                 \n{10}
 LWS  =  [*WSP CRLF] 1*WSP \textit{; linear whitespace}           \n{11}
 SWS  =  [LWS] \textit{; sep whitespace}                          \n{12}
 HCOLON  =  *( SP / HTAB ) ":" SWS                       \n{13}
  \end{alltt}\vspace*{-2em}
  \caption{Extract of the ABNF of the message syntax from the
SIP RFC 3261}\label{fig:rfc3261}
\end{figure}

An ABNF specification consists of a set of derivation rules, each defining
a set of alternatives, separated by \texttt{/}.  An alternative is a
sequence of terminals and nonterminals.  Among the terminals, a quoted
string is case insensitive.  Case sensitive strings must be specified as an
explicit sequence of character codes, as in the {\tt INVITEm} rule
(line~5).  ABNF includes a general form of repetition, $n\mtt{*}m$
\texttt{X}, that indicates that at least $n$ and at most $m$ occurrences of
the terminal or nonterminal \texttt{X} must be present. ABNF also defines
shorthands such as $n\mtt{*}$ for $n\mtt{*}\infty$, $\mtt{*}n$ for
$0\mtt{*}n$, $\mtt{*}$ for $0\mtt{*}\infty$ and $n$ for
$n\mtt{*}n$. Therefore, $1\mtt{*}$\texttt{DIGIT} in the \texttt{CSeq} rule
(line~10) represents a sequence of digits of length at least 1.  Brackets
are used as a shorthand for \texttt{0*1}.

%% Note that unlike the grammars of programming languages, the grammars of
%% network protocol messages rarely if ever contain recursion.  Thus, a
%% certain number of issues that complicate the implementation of parsers for
%% programming language grammars are not a problem in this setting.

\subsection{Hand-writing parsers}
\label{hand}

The specification of the \texttt{CSeq} header in Figure~\ref{fig:rfc3261}
amounts to only four lines of ABNF. However, implementing parsing based on
such an ABNF specification efficiently in a general-purpose language such
as C or C++ often requires many lines of code.  For example, SER and oSIP
contain about 200 and 340 lines of C and C++ code, respectively,
specifically for parsing the \texttt{CSeq} header.  This
\texttt{CSeq}-specific code includes operations for reading individual
characters from the message, operations for transitioning in a state
machine according to the characters that are read, calls to various generic
header parsing operations, and error checking code.  Among the complexities
encountered is the fact that, as shown in Figure~\ref{fig:rfc3261}, a
\texttt{CSeq} header value can stretch over multiple lines if the
continuation line begins with a space or horizontal tab (\texttt{WSP}).

In addition to the constraints described by the ABNF specification, the
parser developer has to take into account constraints on the message
structure that are informally specified in the text of the RFC.  For
example, the \texttt{CSeq} header includes a \textit{CSeq number} expressed
as any sequence of at least one digit (\texttt{1*DIGIT}) and a \textit{CSeq
method} (\texttt{Method}).  The SIP RFC states that the CSeq number must be
an unsigned integer that is less than $2^{31}$ and that the CSeq method
must be the same as the method specified in the request line.  However,
existing hand-written implementations do not always check all these
requirements.  For example, oSIP converts the CSeq number to an integer
without performing any verification.  If the CSeq number contains any
non-numeric characters, the result is a meaningless value.

\subsection{Using parser generators}

PADS and binpac use a type-declaration like format for specifying message
grammars, while GAPA uses a BNF-like format.  Both of these formats require
reorganizing the information in the ABNF specification.  We take PADS as a
concrete example.  Figure~\ref{fig:rfc3261-pads} shows a PADS specification
corresponding to the four lines of ABNF describing the CSeq header.  This
specification is in the spirit of the HTTP specification provided by the
PADS developers~\cite{pads-http}.

\def\nl{$\backslash$}

\begin{figure}[ht]
  \centering\scriptsize
  \begin{alltt}
  bool chkCseqMethod (request_line_t r, Cseq_t c) \{   \n{1}
     return ( r.method == c.method );                  \n{2}
  \}                                                   \n{3}
  
  Ptypedef Puint16_FW(:3:) Cseq_number_t :             \n{5}
     CSeq_t x => \{ 100 <= x \&\& x < 699 \};          \n{6}
  
  Pstruct wsp_crlf_t \{                                \n{8}
     PString_ME(:"({\nl\nl}s|{\nl\nl}t)* {\nl\nl}r{\nl\nl}n":) wsp;            \n{9}  
  \};                                                  \n{10}
  
  POpt wsp_crlf_t o_wsp_crlf_t;          \n{12}
  
  Pstruct lws_t \{                       \n{14}
     o_wsp_crlf_t wsp_crlf;              \n{15}
     PString_ME(:"({\nl\nl}s|{\nl\nl}t)+":) wsp;       \n{16}
  \};                                    \n{17}
  
  POpt lws_t sws_t;                    \n{19} 
  
  Pstruct hcolon_t \{                    \n{21}
     PString_ME(:"({\nl\nl}s|{\nl\nl}t)*":) sp_or_htab;     \n{22}
     ':';   sws_t   sws;               \n{23}  
  \};                                    \n{24}
  
  Pstruct CSeq_t \{                      \n{26}
     PString_ME(:"[Cc][Ss][Ee][Qq]":) name;  \n{27}
     hcolon_t hcolon;                    \n{28}
     CSeq_number_t number;               \n{29}
     lws_t lws;                          \n{30}
     method_t method;                    \n{31}
  \};                                    \n{32}

  Precord Pstruct SIP_msg \{             \n{34}
     request_line_t request_line;        \n{35}
     \textit{[...]}                              
     CSeq_t cseq: checkCSeqMethod(request_line,cseq); \n{37}
     \textit{[...]}                 
  \};                                    \n{39}
  \end{alltt}\vspace*{-2em}
  \caption{PADS specification of the SIP RFC 3261}\label{fig:rfc3261-pads}
\end{figure}

A PADS specification describes both the grammar and the data structures
that will contain the result of parsing the message.  Thus, the rules of
the ABNF specification are translated into what amount to structure
declarations in PADS.  As a PADS structure must be declared before it is
used, the rule ordering is often forced to be different than that of the
ABNF specification.  For example, in the ABNF specification, the
\texttt{CSeq} nonterminal is defined before the \texttt{LWS}, \texttt{SWS},
and \texttt{HCOLON} nonterminals, while in the PADS specification, the
structure corresponding to the \texttt{CSeq} nonterminal is defined
afterwards (line~26).  PADS also does not implement the same default
parsing strategies as ABNF, and thus {\em e.g.}, case insensitive strings
must be specified explicitly using regular expressions (line~27).
Similarly, translating SIP whitespace into PADS requires writing many lines
of specifications (lines 8-19), including regular expressions.  Finally,
the PADS specification must express the various constraints contained in
the RFC text.  Although PADS allows the developer to define constrained
types (lines 5 to 6), which are used here in the case of the CSeq number
(line~29), non-type constraints such as the relationship between the method
mentioned in the request line and the method mentioned in the CSeq header
must be implemented by arbitrary C code (lines 1 to 2 and line~37).

Of these issues, probably the most difficult for the programmer is to
convert ABNF specifications to regular expressions.  Regular expressions
for even simple ABNF specifications are often complex and voluminous.  For
example, a regular expression for a URI has been published that is 45 lines
of code \cite{abnf2regexp}.  While a tool has been developed to convert
an ABNF specification to a regular expression \cite{abnf2regexp}, in the
PADS, GAPA, and binpac specifications that we have seen, the regular
expressions appear to have been written by hand, and sometimes do not
capture all of the constraints specified by the RFC.

\subsection{Integrating a parser with an \newline application}
\label{integrating}

The ease of integrating a parser with an application depends on whether the
parser parses the fields needed by the application, and whether the result
of this parsing is stored in appropriate data structures.  We consider the
issues that arise when using the handwritten oSIP and SER parsers,
and when using a parser generated by a tool such as PADS.

oSIP parses the fixed set of required SIP header fields, and separates the
rest of the message into pairs of a header field name and the corresponding
raw unparsed data.  Applications that do not use all of the information in
the required header fields incur the time cost of parsing this information
and the space cost of storing the result (see Section~\ref{experiments}).
Applications that use the many SIP extensions must parse these header
fields themselves.  The former increases the application time and space
requirements, which can be critical in the case of in-network applications
such as proxies, while the latter leaves the application developer on his
own to develop complex parsing code.

SER provides more fine-grained parsing than oSIP, as it parses only those
header fields that are requested by the application. By default, however,
SER only gives direct access to the top-level subfields of a header, such
as the complete URI.  To extract, {\em e.g.}, only the host portion of a
URI, the programmer must intervene.  One approach is for an application
built using SER to reparse the subfield, to obtain the desired
information.  SER applications are written using a domain-specific language
targeted towards routing, which does not provide string-matching
facilities.  Nevertheless, SER provides an escape from this language,
allowing a SER application to invoke an arbitrary shell script.  A SER
application can thus invoke a script written in a language such as Perl to
extract the desired information.  This approach, however, incurs a high
performance penalty for forking a new process, as we show in Section
\ref{experiments}, and compromises the safety benefits of using the SER
language.  Another approach is to use the SER extension framework, which,
{\em \`a la} Apache \cite{Apache}, allows integrating new modules into the
parsing process.  Although efficient, this approach requires the
programmer to write low level C code that conforms to rather contorted
requirements. Again, incorrect behavior inside a module may compromise the
robustness of the whole application.

Finally, parser generators such as PADS allow the developer to construct
the parser such that it parses only as much of the message as is needed.
However, the generated data structures directly follow the specified
parsing rules, implying that accessing message fields often requires long
chains of structure field references.  Furthermore, all of the parsed data
is stored, which increases the memory footprint.

%% file: robustdev.tex
\section{Robust Parser Development with Zebu}\label{language}

We now present the Zebu language for describing HTTP-like text-based
protocol message formats and related processing constraints.  Zebu is based
on ABNF, as found in RFCs, and extends it with annotations indicating which
message fields should be stored in data structures and other semantic
attributes.  These annotations express both constraints derived from the
protocol RFC and constraints that are specific to the target application.
From a Zebu specification, a compiler automatically generates stubs to be
used by the application to process network messages.

The features of Zebu are driven by the kinds of information that an
application may want to extract from a network protocol message.  We first
consider the features that are needed to do this processing robustly and
efficiently, and then present the corresponding annotations that the
programmer must add to the ABNF specification so that the Zebu compiler can
generate the appropriate stub functions.  Finally, we describe the Zebu
compiler, which performs both verification and code generation, and the
process of constructing an application with Zebu.

\subsection{Issues}
\label{problem}

A HTTP-like text-based network message consists of a command line, a
collection of header fields, and a message body.  The command line
indicates whether the message is a request or a response, and identifies
basic information such as the version of the protocol and the method of a
request message.  A header field specifies a protocol-specific key and an
associated value, which may be composed of a number of subfields.  Finally,
the message body consists of free text whose structure is typically not
specified by the protocol.  Thus, decomposing it further falls out of the
scope of Zebu.

From the contents of a message, an application may need to determine
whether the message is a request or a response, to detect the presence of a
particular header field, or to extract command line or header field
subfields.  Each of these operations involves retrieving a command line or
header field, and potentially accessing its contents.  In a HTTP-like
text-based protocol, each command line or header field normally occupies
one or more complete lines, where each line after the first begins with a
special continuation character.  Thus, as exemplified by the very efficient
SIP parser SER, a parser can be constructed in two levels: a top-level
parser that simply scans each line of the message until it reaches the
desired command line or header field, and a collection of dedicated parsers
that process each type of command line or header field.  The dedicated
parsers must respect both the ABNF specification and any constraints
specified informally in the RFC.  To avoid reparsing already parsed message
elements for each requested parsing operation, the parser should save all
parsed data in data structures for later use, ideally in the format desired
by the application.

This analysis suggests that to enable the Zebu compiler to generate a
useful and efficient parser, the programmer must annotate the ABNF
specification obtained from an RFC with the following information: (1) An
indication of the nonterminal representing the entry point for parsing each
possible command line and header field.  (2) A specification of any
constraints on the message structure that are informally described by the
RFC.  (3) An indication of the message subfields that will be used by the
application.  The first two kinds of annotations are generic to the
protocol, and can thus be reused in generating parsers for multiple
applications.  The third kind of annotation is application-specific.  This
kind of annotation can be viewed as a simplified form of the action that
can be specified when using \texttt{yacc} and other similar parser
generators, in that it allows the programmer to customize the memory layout
used by the parser to the specific needs of the application.

\subsection{Annotating an ABNF specification}

We present the three kinds of annotations required by Zebu, using as an
example an extract of the Zebu specification of a SIP parser, as shown in
Figure \ref{fig:sip3261_zebu}.

\input{sip.fig}

\paragraph*{Parser entry points}

The Zebu programmer annotates the rule for parsing the command line of a
request message with \texttt{requestLine}, the rule for parsing the command
line of a response message with \texttt{statusLine}, and the rules for
parsing each kind of header with \texttt{header}.  Because a command line
or header field cannot contain another command line or header field, the
nonterminals for these lines are no longer useful.  In the case of a
command line, the nonterminal is simply dropped.  Thus, for example, the
ABNF rule for the \texttt{Request-Line} nonterminal
(Figure~\ref{fig:rfc3261}, line 1) is transformed into the following Zebu
rule ({\em c.f.}, Figure~\ref{fig:sip3261_zebu}, line~4):

{\codesize\begin{alltt}
  \textbf{requestLine} \textbf{=} Method SP Request-URI SP SIP-Version
\end{alltt}}

\noindent
In the case of a header field, the description of the key is moved from the
right-hand side of the rule to the left, where it replaces the nonterminal,
resulting in a rule whose structure is suggestive of a key-value pair.  For
example, the ABNF \texttt{CSeq} rule on line 10 of Figure~\ref{fig:rfc3261}
is reorganized into the following Zebu rule ({\em c.f.},
Figure~\ref{fig:sip3261_zebu}, line~38)

{\codesize\begin{alltt}
  \textbf{header} CSeq \textbf{=} 1*DIGIT LWS Method
\end{alltt}}

\noindent
(The delimiter \texttt{HCOLON} is also dropped, as it is a constant of the
protocol).  Some header fields, such as the SIP \texttt{To} header field,
can be represented by any of a set of keys.  In this case, the header is
given a name, which is followed by the ABNF specification of the possible
variants, in braces, as shown in line~44 of Figure \ref{fig:sip3261_zebu}.
As in ABNF, the matching of the header key, and any other string specified
by a Zebu grammar, is case insensitive.

\paragraph*{RFC constraints}

The text of the RFC for a protocol typically indicates how often certain
header fields may appear, whether header fields can be modified, and
various constraints on the values of the header subfields.  The Zebu
programmer must annotate the corresponding ABNF rules with these
constraints.  Constraints are specified in braces at the end of a grammar
rule.  Possible atomic constraints are that a header field is mandatory
(\texttt{mandatory}) and that a header field can appear more than once in a
message (\texttt{multiple}).  For example, in the SIP specification, the header
\texttt{To} is specified to be mandatory and read-only (line~44).  More
complex constraints can be expressed using C-like boolean expressions.  For
example, in Section \ref{hand}, we noted that in a request message, the
method mentioned in the command line must be the same as the method
mentioned in the CSeq header.  This constraint is described in line~7.

Some constraints on header fields are specific to either request or
response messages.  Accordingly, the Zebu programmer must group the request
line and its associated constraints in a {\em request block}, and the
status line and its associated constraints in a {\em response block}.  In
the case of SIP, the request block (lines 2-12) indicates that for the
\texttt{CSeq} header the method must be the same as the method in the
request line (line~7), and that the \texttt{Max-Forwards} header is
mandatory (line~10).  The constraints in the response block (lines 14-18)
have been elided.

\paragraph*{Subfields used by the application}

The parsing functions generated by the Zebu compiler create a data
structure for each command line or header field that is parsed.  By
default, this data structure contains only the type of the command line or
header field and a pointer to its starting point in the message text.  When
the application will use a certain subfield of the command line or message
header, the Zebu programmer can annotate the nonterminal deriving this
subfield with an identifier name.  This annotation causes the Zebu compiler
to create a corresponding entry in the enclosing command line or header
field data structure.  For example, in line~4, the Zebu programmer has
indicated that the application needs to use the method in the command
line, which is given the name \textsf{method}, and the URI, which is given
the name \textsf{uri}.

By default, a subfield is just represented as a pointer to the start of its
value in the message text.  This is the case of \textsf{method} and
\textsf{uri} in our example.  Often, however, the application will need to
use the value in some other form, such as an integer.  The Zebu programmer
can additionally specify a type for a named value, either at the
nonterminal reference or at its definition.  For example, in line~38 the
CSeq number is specified as being a {\sf uint32}.  Nonterminals can also be
specified as structures (\texttt{struct}), unions (\texttt{union}), and
enumerations (\texttt{enum}).  A structure collects all derived named
subfields.  As illustrated in the case of \texttt{Request-URI} (line~25), a
structure may even be used in the case of an alternation, when the
application does not need to know from what element of the alternation a
named entry is derived.  A union, in contrast, records which alternation
was matched and in each case only includes subfields derived from the given
alternation.  Finally, an enumeration is a special case of union in which
the only information that is recorded is the identity of the matching
alternation; the matched data is not stored.  In line~20, for example,
\texttt{Method} is specified as being an enumeration, because the
application only needs to know whether the method of the message is one of
the standard ones or an extension method, but does not need to know the
identity of the extension method in the latter case.

An application may use the information in certain subfields only in
some exceptional cases.  The Zebu constraint \fname{lazy} allows the
programmer to specify that a specific subfield should not be parsed until
requested by the application.  For example, in the SIP specification,
\fname{Request-URI} has this annotation (line~25).

\subsection{The Zebu compiler}

The Zebu compiler verifies the consistency of the ABNF specification and
the annotations added by the programmer, and then generates stub functions
allowing an application to parse the command line and header fields and
access information about the parsed data.  The Zebu compiler is around
3700 lines of OCaml code.  A run-time environment defining various utility
functions is also provided, and amounts to around 700 lines of C code.

\paragraph*{Verifications}

Although RFCs are widely published and form the {\em de facto} standard for
many protocols, we have found some errors in RFC ABNF specifications.
These are simple errors, such as typographical errors, but still they
complicate the process of translating an ABNF specification into code,
whether done by hand or using a parser generator.  The Zebu compiler thus
checks basic consistency properties of the ABNF specification: that there
is no omission ({\em i.e.}, each referenced rule is defined), that there is
no double definition, and that there are no cycles.

Additionally, the annotations provided by the Zebu programmer must be
consistent with the ABNF specification.  For example, in line~30, the
nonterminal \fname{Global-Failure} is annotated with \texttt{uint16}.
This non-terminal is specified to be an alternation of strings, and thus
the Zebu compiler checks that each element of this alternation represents
an unsigned integer that is less than $2^{16}$.

\paragraph*{Code generation}

An application does not use the data structures declared in a Zebu
specification directly, but instead uses stub functions generated by the
Zebu compiler.  The use of stub functions allows parsing to be carried out
lazily, so that only as much data is parsed as is needed to fulfill the
request of a given stub function call.  As illustrated in Figure
\ref{fig:stubs}a, stub functions are generated for determining the type of
a message (request or response), for parsing the command line and the
various headers, for accessing individual header subfields, and for
managing the parsing of subfields designated as \texttt{lazy}. The names of
these stub functions depend on the specific structure of the grammar, but
follow a well-defined schema that facilitates their use by the application
developer.

\input{app.fig}

The parsing functions generated by the Zebu compiler use the two-level
parsing strategy described in Section~\ref{problem}.  Header-specific
parsers use the PCRE~\cite{pcre} library for matching the regular
expression of a header value that has been derived from the ABNF
specification.  The parsing functions contain run-time assertions that
check the constraints specified in the RFC.  Once a header is parsed and
checked, its named subfields, if any, are converted to the specified types
and stored in the data structure associated with the header.  The values of
the named subfields can then be accessed using the ``get'' stub functions.

\subsection{Developing an application with Zebu}\label{sec:zebu_app}

The developer defines the application logic as an ordinary C program, using
the stub functions to access information about the message content.
Figure~\ref{fig:stubs}b illustrates the implementation of an application
that extracts the host information from the URI stored in the \fname{From}
header field of an \fname{INVITE} message.  This kind of operation is
useful in, {\em e.g.} an intrusion detection system, which searches
for certain patterns of information in network messages.

The application uses the stubs generated from the SIP message grammar
specification to access the required information.  The application
initially uses the functions
\texttt{sip3261\_\-Method\_\-getType} and
\texttt{sip3261\_\-RequestLine\_\-getMethod} to determine whether the
current message is an \texttt{INVITE} request (line~6).  If so, it uses the
function \texttt{sip3261\_\-parse\_\-headers} to parse the \texttt{From}
header field (line~8), and then the functions
\texttt{sip3261\_\-header\_\-From\_\-getUri} and
\texttt{sip3261\_\-get\_\-header\_\-From} to extract the URI (line~9).
Line 46 of the Zebu SIP specification indicates that the parsing of the URI
should be lazy, so the function
\texttt{sip3261\_\-Lazy\_\-Addr\_\-spec\_\-getParsed} is used to force the
parsing of this subfield (line~10).  After a check that the host name is
present (line~13), its value is extracted using the function
\texttt{sip3261\_\-Option\_\-Str\_\-getVal} in line~14.

Overall, due to the annotations in the Zebu specification, stub functions
are available to access exactly the message fragments needed by the
application.  Similarly, memory usage is limited to the application's
declared needs.

%% file: experiments.tex
\section{Experiments}\label{experiments}

A robust network application must accept valid messages, to provide
continuous service, and reject invalid network messages, to avoid
corrupting its internal state.  As the parser is the front-line in the
treatment of network messages, it has a key role to play in providing this
robustness.  In this section, we evaluate the robustness improvement
offered by Zebu, by comparing the reaction of Zebu-based parsers and a
variety of existing parsers to valid and invalid network messages.  Our
experiments are based on a mutation analysis technique.

For SIP, we compare with the oSIP and SER parsers previously described in
Section \ref{issues}.  For RTSP, we use the parser in the widely used VLC
media player and streaming server \cite{vlc}, and the parser provided by
the LiveMedia library~\footnote{LiveMedia: Streaming Media,
\scriptsize{\fname{http://www.livemediacast.net/}}}.
Figure~\ref{fig:sizes} shows the sizes of the ABNF and Zebu specifications
of the message grammars for SIP and RTSP.  The Zebu specification is
longer, because it includes rules that are mentioned only by reference to
another RFC in the original SIP and RTSP specifications.  Figure
\ref{fig:sizes} also shows the number of lines of code in the oSIP, SER,
VLC, and LiveMedia parser implementations.

\input{sizes.fig}

In the rest of this section, we first introduce mutation analysis, and
then compare the robustness of existing SIP and RTSP parsers with that
of the corresponding Zebu-based parsers.  Finally, we evaluate the
performance of Zebu, showing that Zebu-based parsers are often as
efficient as hand-written ones.

\subsection{Robustness evaluation}

Mutation analysis is a fault-based testing technique for unit-level testing
\cite{DeMillo:computer1978}.  Traditional mutation testing involves
introducing small changes, {\em i.e.}, mutations, in program source code,
to determine whether a given test suite is sufficient to distinguish
between correct and incorrect programs. In our case, however, we are
interested in assessing the robustness of the program, {\em i.e.}, the
parser, and thus we introduce mutations into the test data, {\em i.e.}, the
network messages, rather than into the program source code.  We use
mutation rules both to generate invalid messages and to generate valid
messages that have properties that are known to be challenging for network
protocol message parsers.  A robust parser should reject the invalid
messages and accept the valid ones.

To generate invalid messages, we have defined a set of mutation rules for
messages based on ABNF structure:

\begin{itemize}
\item Mutations on the characters set. Message literals are derived from a
  fixed set of possible characters. The first, middle, or last character of
  a message literal is replaced with any character outside the valid set.

\item Mutations on repetitions.  As described in Section~\ref{issues_abnf},
  ABNF offers a generic mechanism of repetition. Mutants are chosen to
  describe an invalid number of repetitions.

\item Mutations based on constraints. Protocol specifications include
  additional constraints not specified in the message grammar about the
  values of header subfields. For example, the response code of a SIP
  response is not only an unsigned integer of three digits, but its value
  must also be less than 699 (see Figure~\ref{fig:sip3261_zebu}).  Mutants
  are chosen that violate these constraints.
\end{itemize}

To generate valid but problematic messages, we have extended our character
set mutation rule to create messages of the form suggested by the SIP
Torture Test Message RFC~\cite{rfc4475}.  This RFC describes a set of valid
SIP messages that test corner cases in SIP implementations.

To compare the robustness of Zebu-based applications to applications based
on hand-crafted parsers, we consider the parsing of the principal fields of
a network protocol message.  For SIP, these fields are the command line and
the six mandatory header fields, while for RTSP they are the command line
and the header fields \fname{Transport}, \fname{CSeq} and
\fname{UserAgent}.  We drive each of the parsers listed in
Figure~\ref{fig:sizes} using minimal applications that request access to
these fields.  The Zebu-based applications \fname{log-Zebu-SIP} and
\fname{log-Zebu-RTSP}, for SIP and RTSP respectively, consist of a few
lines of C code that log statistical information about incoming
messages. These applications use the stubs generated by the Zebu compiler
to access network messages, analogous to the code illustrated in
Figure~\ref{fig:stubs}b.  The SER application, \fname{log-SER} is written
using the SER configuration language to access the information in the
various fields.  The other applications, \fname{log-oSIP} using oSIP,
\fname{log-VLC} using VLC, and \fname{log-LiveMedia} using LiveMedia, are
written in C using the appropriate API functions provided by the given
parser.

\paragraph*{Invalid messages}
In our first set of tests, we apply our mutation rules to SIP and RTSP
messages, generating a stream of invalid messages, which we then send to
each of the SIP and RTSP applications, respectively.  As shown in
Figure~\ref{fig:mutations}, while the Zebu-based applications detect every
mutant as representing an invalid message, none of the hand-crafted parsers
detects more than about 25\% of the injected mutants. This situation may
have a critical impact. In the case of SIP for example, we have crashed SER
via a stream of 2416 incorrect messages, sent within of 17 seconds.
Because SER is widely used for telephony, which is a critical service, the
ability to crash the server is unacceptable.

\input{mutations.fig}

\vspace*{-2em}\paragraph*{Valid messages}
While message parsers should detect erroneous messages as early as possible
to preserve the robustness of the applications that use them, they also
must correctly parse valid messages. The SIP Torture Test Message
RFC~\cite{rfc4475} describes a set of valid SIP messages that test corner
cases in SIP implementations.  Guided by this RFC, we have extended our
character set mutation rule to generate mutants that are valid SIP messages
but are designed to torture a SIP implementation.
Figure~\ref{fig:mutations_valid} shows that up to about 4\% of the valid
messages are rejected by hand-crafted SIP parsers. By comparison, the
Zebu-based SIP parser strictly follows the message grammar.

\input{mutations_valid.fig}

We have tried an analogous experiment with the RTSP applications, but the
VLC and LiveMedia parsers are quite lax in their parsing of the message
elements, such as the URI, that are covered by the SIP Torture Test RFC,
and thus all three applications accept all of the mutated messages.

\subsection{Performance Evaluation}

We now compare the performance of Zebu-based parsers to that of
hand-crafted ones. Our results are only for SIP, which is the most
demanding in terms of performance. For our experiments, we have implemented
four versions of the SIP message statistics reporting application described
in Section~\ref{sec:zebu_app}. In each case, the application records the
host information of the URI stored in the \fname{From} header field of an
\fname{INVITE} message.  The first version (\fname{inv-SER-module}) is
implemented as a dedicated SER module to obtain full access to the internal
data structures of SER.  The second version (\fname{inv-SER-exec}) is
written using the configuration language of SER and relies on the escape
mechanism provided by SER to invoke \fname{sed} to extract the host
information, as described in Section \ref{integrating}.  The third version
(\fname{inv-oSIP}) is implemented using a few lines of C code on top of the
oSIP SIP stack.  The last version (\fname{inv-Zebu}) is the Zebu-based
application depicted in Figure~\ref{fig:stubs}b.

Our application illustrates the case where an application such as a
intrusion detection system needs to access a fragment of a header subfield.
To explore the effect that various kinds of messages have on the parsing
performance for such an application, we consider a collection of
\fname{INVITE} messages, which are relevant to our application, and an
example of a non-\fname{INVITE} message, which is not.  Among the
\fname{INVITE} messages, in \fname{INVITE}$_1$ the \fname{From} header
field contains only the URI subfield and an required tag subfield; all of
the other subfields, which are optional, are omitted.  This entails the
minimal processing for a message that is relevant to the application.  The
remaining \fname{INVITE} messages, \fname{INVITE}$_2$ and
\fname{INVITE}$_3$, show the effect of varying the position of the
\fname{From} header field.  In \fname{INVITE}$_2$, the \fname{From} header
field is the first of 34 header fields, while in \fname{INVITE}$_2$ it is
the last of 34 header fields.  The non-\fname{INVITE} message is a
\fname{BYE} and has 7 headers.

\input{performance.fig}

Our experiments were performed using a Pentium III (1GHz) as the server,
which is stressed by a bi-processor Xeon 3.2Ghz client.
Figure~\ref{fig:SIP_bench} compares the parsing time for each of the
applications to that of \fname{SER-module}, which has the fastest parser
among the existing parsers that we tested.  

SER uses the efficient two-level parsing strategy described in Section
\ref{problem}, to parse only the header fields that are relevant to
the application.  The parsing done by \fname{inv-SER-module} is
particularly efficient in the case of \texttt{INVITE} messages, as the
information required by the application is already available in the
SER internal data structures.  The parsing done by
\fname{inv-SER-exec} is roughly as efficient as that done by
\fname{inv-SER-module} for the non-\fname{INVITE} message.  The parsing
done by \fname{inv-SER-exec} for the \fname{INVITE} messages, on the
other hand, is up to 648 times slower, because it forks a \fname{sed}
process.  Despite the bad performance in this case, the use of the
configuration language of SER remains relevant, because it provides
ease of programming and safety, which are not provided by the use of a
SER module.

The parsing done by \fname{inv-oSIP} is over 13 times slower than the
parsing done by \fname{inv-SER-module} for \fname{INVITE} messages and
over 10 times slower for the non-\fname{INVITE} message.  In both
cases, oSIP parses the six required SIP headers (plus two more
required headers in the case of a \fname{REGISTER} message) and stores
pointers to the starting point of each sub-field.  As the application
requests information about the \fname{INVITE} header field, oSIP
additionally copies the subfields into a data structure that is
provided to the application, roughly doubling the execution time.  No
results are presented for \fname{inv-oSIP} for \fname{INVITE}$_3$,
because oSIP crashes on this message.

Finally, while Zebu follows the same two-level parsing strategy as SER, the
parsing done by \fname{inv-Zebu} is significantly slower than the parsing
done by \fname{inv-SER-module} for the \fname{INVITE} messages, because
Zebu checks the URI more rigorously than SER.  On the other hand, Zebu is
significantly more efficient than SER for the non-\fname{INVITE} messages.
SER is directed towards routing applications, and thus it
always parses the \fname{Via} header, which is essential in the routing
process, although irrelevant to our application.  Thus, Zebu provides
better performance in such cases by being more closely tailored to the
needs of the application, and retains safety, which is lost in SER when
using the module approach.

%% file: sizes.fig
%SIP ABNF = ~700 lines, RTSP ABNF = 214 lines

%% SER dans un repertoire e part:11982 lignes de code
%% OSIP dans un repertoire e part : 13277 lignes de code

%% VLC distribue un peu partout : Environ 1200 lignes
%% LIVEMEDIA RTSP Library un peu partout : Environ 1000 lignes
%% Mais comme c'est du code un peu distribue, c'est un peu difficile...

\begin{figure}[th]\scriptsize
  \centering
\begin{tabular}{|c|c|c|c|c|}
\hline
Protocol & ABNF size & Zebu spec size &    Parser    & Parser size  \\\hline
\textbf{SIP}      & 700 (approx) & 1081 &    oSIP      & 11982        \\\cline{4-5}
         &           &    &SER       & 13277        \\\hline
\textbf{RTSP}     & 200 (approx) & 330 &    VLC       & 1200 (approx)\\\cline{4-5}
         &           &    &LiveMedia & 1000 (approx)\\\hline
\end{tabular}   
  \caption{The sizes of the SIP and RTSP message grammars, and the sizes of
  existing parsers.  Sizes in lines of code.}\label{fig:sizes}
\end{figure}

%% file: performance.fig
\def\head#1{\multicolumn{2}{|c|}{\texttt{#1}}}

\begin{figure*}[ht]\scriptsize
  \centering  
  \begin{tabular}{|c|c|c|c|c|c|c|c|c|c|}
    \cline{3-10} 

    \multicolumn{2}{c|}{} &
    \head{inv-SER-module} &
    \head{inv-SER-exec} &
    \head{inv-oSIP} & 
    \head{inv-Zebu} \\
	
    \cline{2-10}
    
    \multicolumn{1}{c|}{} &
    \textbf{Message size} &
    \textbf{Cycles} & \textbf{Ratio} & 
    \textbf{Cycles} & \textbf{Ratio} & 
    \textbf{Cycles} & \textbf{Ratio} & 
    \textbf{Cycles} & \textbf{Ratio} \\

  \hline

    INVITE$_1$ & 
     697 & 
     13 788    & 1 & 
     7 593 550 & 551 & 
     182 703   & 13  & 
     51 054    & 4    \\
    \hline
    
    INVITE$_2$ & 
     1 734 & 
     13 595    & 1  & 
     8 803 456 & 648 & 
     276 275   & 20  & 
     80 270    & 6    \\
     \hline

    INVITE$_3$ & 
     1 734 &
     32 045    & 1 & 
     10 015 827 & 313 & 
     -         & - & 
     133 164   & 4    \\
    \hline

%     ACK        &
%      330 &
%      16 652  & 1 & 
%      36 799  & 2 & 
%      104 761 & 6 & 
%      6 089   & 0.3  \\
%     \hline

    BYE        & 
     334 &
     10 252  & 1 & 
     10 765  & 1 & 
     105 773 & 10 & 
     6 037   & 0.6  \\
    \hline

%     REGISTER   & 
%      570 &
%      15 224  & 1   & 
%      15 314  & 1   &  
%      161 358 & 11 & 
%      6 310   & 0.4    \\
%     \hline
    
  \end{tabular}
  \caption{Performance of SIP applications (time in
  cycles, ratio as compared to \texttt{inv-SER-module})}\label{fig:SIP_bench}
\end{figure*}

%% file: related.tex
\section{Related Work}\label{rel}

Parser generators such as DATASCRIPT~\cite{Back:gpce02},
PacketTypes~\cite{DBLP:conf/sigcomm/McCannC00},
PADS~\cite{DBLP:conf/pldi/FisherG05}, GAPA~\cite{gapa} and
binpac~\cite{Pang:im06} have been recently developed to address the growing
complexity of network protocol messages. However, as described in
Section~\ref{issues}, these tools do not fulfill all the requirements of
network application developers.  APG \cite{apg} is a parser generator that
accepts ABNF directly.  Semantic actions are specified via callback
functions rather than annotations on the grammar.  We have found the use of
such callback functions to be somewhat heavyweight, in our experience in
using APG.  Furthermore, APG is not specific to HTTP-like text-based
protocols, and thus cannot implement the two-level parsing strategy
outlined in Section \ref{problem}, which we have found (Section
\ref{experiments}) essential to obtaining good performance.

Domain-specific languages have been used successfully in various
application domains including operating
systems~\cite{DBLP:conf/sosp/LooCHMRS05,Merillon-al:osdi2000} and
networks~\cite{DBLP:conf/pldi/GayLBWBC03,Hicks-al:icfp98}.  Several of
these languages have explicitly targeted improving system robustness.  The
Devil language, in the domain of device-driver development, provides
high-level abstractions for specifying the code for interacting with the
device, and performs a number of compile-time and (optional) run-time
verifications to check that the specifications are
consistent~\cite{Reveillere-Muller:dsn2001}.  The language Promela++ for
specifying network protocols, can be translated automatically both into the
model checking language Promela \cite{holzmann:91} and into efficient C
code, thus easing the development of a protocol implementation that is
both verified and efficient \cite{Basu:98}.

Mutation analysis has been used to test the robustness of other software
components, such as operating systems \cite{Duraes:02}, network intrusion
detection systems \cite{Vigna:04}, and databases \cite{Tuya}.  Our work is
most similar to the work on network intrusion detection systems, which
also mutates network protocol messages.

%% file: conclusion.tex
\section{Conclusion}
\label{conclusion}

In this paper, we have presented the Zebu declarative language for
describing protocol message formats and related processing
constraints. Zebu builds on the ABNF notations typically used in RFCs
to describe protocol grammars.  In evaluating Zebu, we have
particularly focused on analyzing the improvement in robustness that
it provides.  For this, we have defined a test methodology based on a
mutation analysis that injects errors into network messages. We have
applied our test methodology to SIP and RTSP servers by comparing
existing parsers with Zebu-generated ones.

The results of our experiments show that nearly 4 times more erroneous
messages are detected by the Zebu-based parser than by widely-used
hand-written parsers. In the case of SIP, we were able to crash the widely
used SER parser~\cite{kuthan:ser-03} via a stream of 2416 incorrect
messages, sent within a space of 17 seconds.  Because SER is used for
telephony, which is a critical service, the ability to crash the server is
unacceptable.  We have also found valid messages that are not accepted by
the SER and oSIP parsers, which can similarly have a critical impact.
Finally, we have shown that the added safety and robustness provided by
Zebu does not significantly impact performance.  In the case of SIP, in
micro-benchmarks, we have found that a Zebu-based parser is often as
efficient as a hand-crafted one.